\documentclass[twocolumn,showpacs,preprintnumbers,amsmath,amssymb,floatfix,prl,superscriptaddress]{revtex4-1}
\usepackage[utf8]{inputenc}
\usepackage{setspace}
\usepackage{graphicx}
\usepackage{graphics}
\usepackage{subfigure}
\usepackage{dcolumn}
\usepackage{bm}
\usepackage{color}
\usepackage{SIunits}
\usepackage[normalem]{ulem}
\usepackage[francais]{babel}

\bibpunct{[}{]}{,}{n}{,}{,}

\begin{document}
\title{Plasma mirrors as a path to the Schwinger limit}

\author{L. Chopineau}
\affiliation{Université Paris-Saclay, CEA, CNRS, LIDYL, 91191, Gif-sur-Yvette, France\\}
\author{A. Denoeud} 
\affiliation{Université Paris-Saclay, CEA, CNRS, LIDYL, 91191, Gif-sur-Yvette, France\\}
\author{A. Leblanc} 
\affiliation{LOA, ENSTA ParisTech, CNRS, Ecole polytechnique, Université Paris-Saclay, 828 bd des Maréchaux, 91762 Palaiseau cedex France\\}
\author{E. Porat} 
\affiliation{The School of Physics and Astronomy, Tel Aviv University, Tel Aviv 69978, Israel}
\affiliation{Applied Physics department, Soreq NRC, Yavne 81800, Israel\\}
\author{Ph. Martin} 
\author{H. Vincenti} 
\author{F.Qu\'er\'e}
\affiliation{Université Paris-Saclay, CEA, CNRS, LIDYL, 91191, Gif-sur-Yvette, France\\}

\date{\today}
\newcommand{\e}{\mbox{\boldmath$\eta$}}
\newcommand{\x}{\mbox{\boldmath$x$}}
\newcommand{\si}{\mbox{\boldmath$\xi$}}

\begin{abstract}
Reaching light intensities above $10^{25}$ W/cm$^{2}$ and up to the Schwinger limit ($10^{29}$ W/cm$^{2}$) would enable testing decades-old fundamental predictions of Quantum Electrodynamics. A promising yet challenging approach to achieve such extreme fields consists in reflecting a high-power femtosecond laser pulse off a curved relativistic mirror. This enhances the intensity of the reflected beam by simultaneously compressing it in time down to the attosecond range, and focusing it to sub-micron focal spots. 
Here we show that such curved relativistic mirrors can be produced when an ultra-intense laser pulse ionizes a solid target and creates a dense plasma that specularly reflects the incident light. This is evidenced by measuring for the first time the temporal and spatial effects induced on the reflected beam by this so-called 'plasma mirror'. 
The all-optical measurement technique demonstrated here will be instrumental for the use of relativistic plasma mirrors with the emerging generation of Petawatt lasers, which constitutes a viable  experimental path to the Schwinger limit.  

\end{abstract}

\maketitle

Quantum field theory predicts that even the most perfect vacuum has a complex structure, characterized by a jostling of so-called virtual particles. Probing the non-linear optical properties of vacuum, resulting from the coupling of light with these virtual particles, would enable unprecedented tests of these predictions \cite{marklund2006nonlinear,di2012extremely,mourou2006optics}. The typical amplitude of the electromagnetic fields required to do so corresponds to the onset of the Sauter-Schwinger effect \cite{sauter1931verhalten, Heisenberg1935, schwinger1951gauge}, where charged particle-antiparticle pairs are predicted to spontaneously appear from a vacuum in which a sufficiently strong electric field is applied. For electron-positron pairs, this is expected for field amplitudes exceeding $E_s=m_e^2c^3/e \hbar=1.32 \times 10^{18} V/m$, corresponding to electromagnetic waves with an intensity $I_s \geq c\epsilon_0 E_s^2/2 \simeq 4.7 \times 10^{29}$ $W/cm^2$. Due to these considerable values, this so-called Schwinger limit has never been reached or even approached in experiments \cite{ringwald2001pair}.

The concentration of light energy allowed by ultrashort lasers \cite{strickland1985compression,bahk2004generation} has raised the hope of approaching such intensities by focusing near-visible laser light \cite{gerstner2007laser}. The critical value $I_s$ however still far exceeds the present record of laser intensity of $\simeq 5.10^{22}$ $W/cm^2$, recently achieved with a tightly-focused Petawatt (1 PW=$10^{15}$ W) femtosecond (1 fs=$10^{-15}$ s) laser \cite{Yoon19}. Further increasing the laser pulse energy appears as a technologically hopeless path to close this gap of about seven orders of magnitude. A more realistic approach would consist in considerably increasing the concentration of this light energy. This requires a conversion to electromagnetic waves of shorter wavelengths, which can be more tightly focused in space and compressed in time.

A promising path to implement such a frequency conversion for high-power laser pulses is the concept of \textsl{curved relativistic mirror} \cite{landecker1952possibility,bulanov2003light}. Upon  reflection on a mirror moving at $v \lesssim c$, an ultraintense laser pulse gets compressed in time and down-converted in wavelength by a factor $\simeq 4 \gamma^2$ (with $\gamma=1/\sqrt{1-v^2/c^2}$ the mirror's Lorentz factor) due to the Doppler effect. For large $\gamma$, the reflected pulse can be compressed down to the attosecond (1 as=$10^{-18}$ s) time range, and can now converge to a sub-micron focal spot, thus boosting the intensity of the initial laser by orders of magnitude. Although appealing, the major and so-far unsolved challenge of this approach is the experimental generation of such curved relativistic mirrors. 

\begin{figure*}[t]
\centering \includegraphics[width=\linewidth]{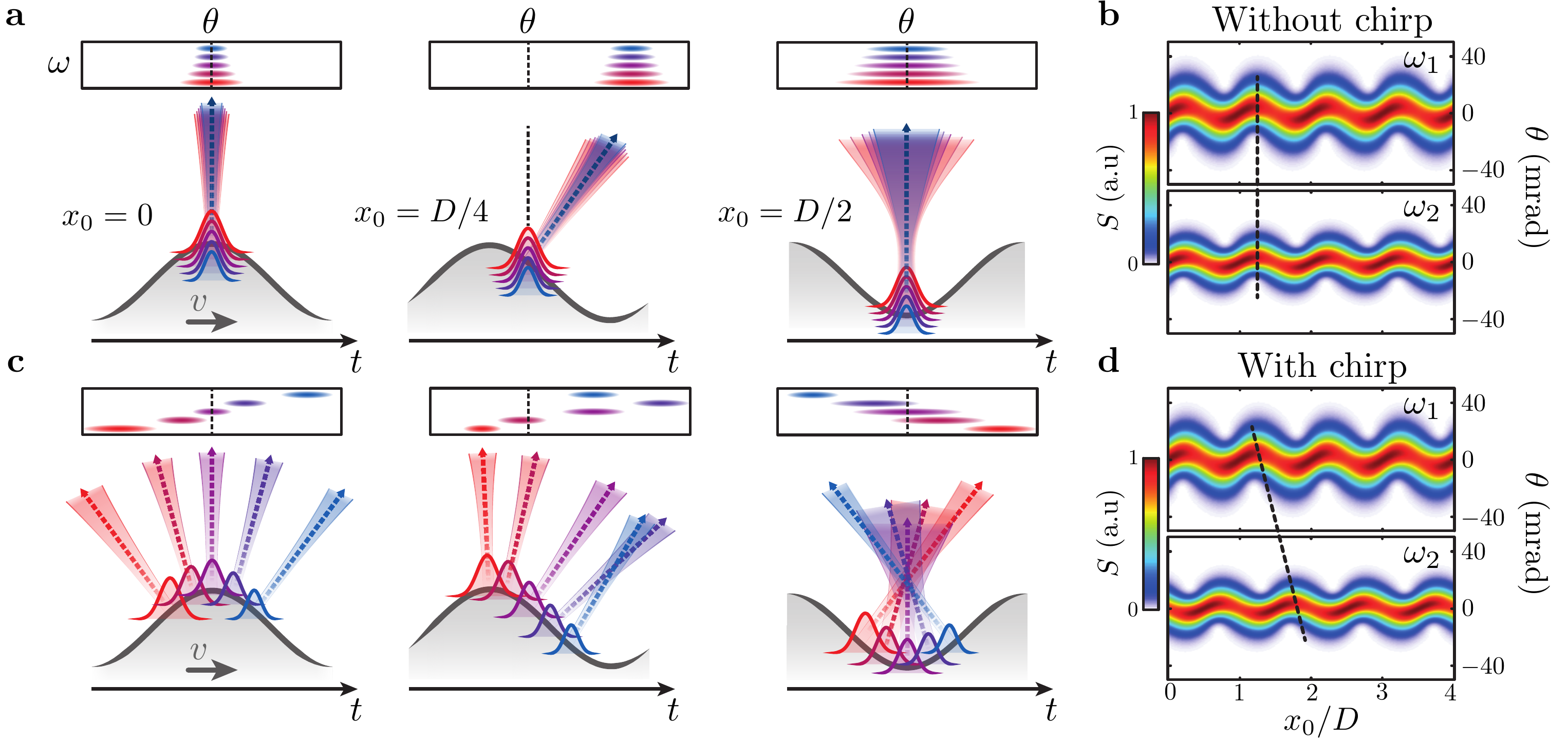}
\vskip -0.25cm 
\caption{\textbf{Principle of dynamical ptychography}. This figure presents an example of dynamical ptychography, in a case where the diffracting object (grey area in a and c) is a sinusoidal reflective surface. The ptychographic measurement consists in measuring the diffraction pattern (upper insets in a and c) of the illuminating beam (Gaussian profiles in a and c) while scanning the position offset $x_0$ of this surface with respect to the beam (from left to right in panels a-c). In dynamical ptychography, the probe results from the superposition of multiple frequencies $\omega$ represented by the different colored Gaussian profiles. Ptychographic traces $S(\theta,x_0,\omega)$ are measured for several of these frequencies, and such traces are displayed in panels b and d for two of these frequencies $\omega_1$ and $\omega_2$. In this example, the object is assumed to drift at constant velocity $v$. The temporal information on the probe pulse is encoded in the frequency dependence of the phase of the oscillating ptychographic traces, as illustrated here for the case of an optimally-compressed probe (panels a-b) and a chirped probe (panels c-d). The black dashed lines in b-d are guides for the eye. 
} 
\vskip -0.5cm
\label{fig1}
\end{figure*}

Different implementations have been proposed \cite{bulanov2003light, gordienko2005coherent,gonoskov2011ultrarelativistic,PhysRevLett.123.105001} and debated \cite{solodov2006limits} to achieve this goal. Experimentally, significant advances have already been made in the last 15 years, using so-called plasma mirrors: these are dense plasmas produced at the surface of initially-solid targets, ionized by intense fs laser pulses \cite{kapteyn1991prepulse,thaury2007plasma}. They have the ability to specularly reflect high-power ultrashort laser pulses, just like ordinary mirrors do for ordinary light. 
At intensities exceeding $10^{18}$ $W/cm^2$, the laser-driven oscillation of the plasma surface becomes relativistic, and thus induces a periodic Doppler effect on the reflected beam \cite{gonoskov2011ultrarelativistic, lichters1996short,thaury2010high, baeva2006theory, dromey2006high}. This results in the generation of a train of attosecond light pulses -with one pulse every laser period- associated in the frequency domain to a comb of harmonics of the laser frequency. In addition to these temporal effects, the surface of plasma mirrors can get strongly curved under the effect of the radiation pressure exerted by the incident laser field, which can provide a way to focus these attosecond pulses right after their generation. Implemented with the emerging generation of PW lasers, this scheme appears as one of the few viable paths towards the Schwinger limit \cite{PhysRevLett.123.105001}. 


Pursuing this path requires the accurate measurement of the spatio-temporal properties of the reflected beam, down to the attosecond scale in time and nanometric scale in space, to be able to both assess and optimize the properties of the relativistic mirror. Despite the development of attosecond metrology in the last 20 years \cite{krausz2009attosecond,quere2005temporal,orfanos2019attosecond}, attosecond pulses generated from plasma mirrors have never been accurately characterized in time, due to the challenge of implementing such advanced techniques in the extremely harsh conditions of laser-plasma experiments \cite{nomura2009attosecond, quere2009attosecond}. In this article, we report the first spatio-temporal characterization of attosecond pulses generated from relativistic plasma mirrors with a 100 TW-class fs laser, using an all-optical technique. We thus get direct evidence for the effects that compress the light energy of the reflected beam in time and space. The measurement method that we demonstrate can be realistically scaled to the most powerful fs laser systems available to date, and thus constitutes a milestone in the quest for the Schwinger limit. 


\section*{Principle of dynamical ptychography}

The measurement method implemented in our experiment is an extension of a powerful lensless imaging technique known as ptychography \cite{rodenburg2008ptychography}. This technique consists in illuminating a microscopic object, described by a transmission or reflection function $O(x-x_0)$, with a focused beam of coherent light of wavevector $k_0$, described by a field $E(x)$. The angular diffraction pattern $S(\theta,x_0)\propto \left|\int dx \ O(x-x_0)E(x)e^{i \sin\theta k_0 x}\right|^2$ produced as the beam diverges away from the object is measured as a function of the position $x_0$ of the object with respect to the beam. As an example, we consider the simple case of a sinusoidual surface as an object Fig.1a. This object is illuminated by a probe beam whose size at focus is smaller than the surface oscillation period $D$. The resulting  ptychographic trace $S(\theta,x_0)$ is displayed in Fig.1b. As the  position $x_0$ is scanned, the peaks and dips of the surface are alternatively probed, and both the angular width and direction of the diffracted beam oscillate.
The power of ptychography lies in the fact that iterative phase retrieval algorithms can be applied to such a trace, to reconstruct the spatial structures $O(x)$ of the object and $E(x)$ of the illuminating beam, in amplitude and phase. Ptychography is thus an advanced spatial measurement method for both microscopic objects and coherent beams of light or particles \cite{thibault2008high}.

We now consider an illuminating probe consisting of an ultrashort pulse of light, and aim at adding temporal resolution to ptychography, in order to obtain both spatial and temporal information on this beam. We show that this is possible by using an object $O(x,t)$ which evolves in time in a known manner. The measurement then consists in spectrally-resolving the diffraction pattern of the probe beam on this evolving object as the offset position $x_0$ is scanned. This provides a 3D dataset $S(\theta,x_0,\omega)$, where $\omega$ is frequency within the spectral width of the probe pulse. To illustrate this measurement scheme, we now assume that the object of Fig.1 drifts in time with a known and constant velocity $v$ (Fig.1).

An ultrashort pulse of light is formed by the superposition of multiple frequencies. For a given spectrum, the shortest pulse is formed if all frequencies are all perfectly synchronized in time (i.e., group delay dispersion is zero). In such a case, all these frequencies will probe the evolving object at the same instant (Fig.1a). The multiple ptychographic traces measured at different frequencies $\omega$ would then be observed to oscillate in phase (Fig.1b).

We now consider a probe pulse with the same spectrum, but with the frequencies arriving at different times \footnote{The arrival time of frequency $\omega$ here refers to the group delay at this frequency, i.e. it corresponds to the temporal position of the pulse formed by the superposition of frequencies within a small interval centered at $\omega$.}. This corresponds to a chirped pulse whose duration is longer than the minimum allowed by the pulse spectral bandwidth. Different frequencies $\omega$ within the pulsed beam will now probe the object at different instants of its motion, and will get diffracted differently (Fig.1c). The oscillations of the multiple ptychographic traces measured at different frequencies would then be observed to be dephased (Fig.1d), i.e. to present a maximal deflection for different values $x_m(\omega)$ of $x_0$. If the object velocity $v$ is known, this dephasing directly encodes the chirp of the probe pulse, through the following straightforward relationship:
\begin{equation}
\tau(\omega)=x_m(\omega)/v
\end{equation}
where  $\tau(\omega)$ is the group delay of frequency $\omega$ within the probe spectrum. Combined with the spatial information provided by each ptychographic trace, we can thus get access to the complete spatio-temporal structure of the illuminating beam. 


This measurement scheme, which we call dynamical ptychography, is very general and can in principle be applied to very different types of objects and probes, over a broad range of time scales, to determine the temporal properties of the probe if the evolution of the object is known, or vice versa. 

\section*{Application to attosecond pulses \\ from plasma mirrors}

Here we apply this general measurement scheme to determine the spatio-temporal structure of attosecond pulses generated from plasma mirrors. This calls for a diffracting object evolving very quickly in time, i.e. typically moving by a fraction of its spatial period on an attosecond time scale. We fulfill this condition by using a \textit{transient optical grating} as the diffracting structure: this consists in an evolving spatial interference pattern, applied on the ultraintense laser beam that drives the interaction with the plasma mirror and generates the attosecond pulses. The spatio-temporal structure of this driving field is imprinted on the dynamics of the plasma mirror and hence on the generated attosecond pulses. This acts as the evolving diffracting object for the dynamical ptychographic measurement of this light source (see SM). 

\begin{figure*}[t]
\centering \includegraphics[width=\linewidth]{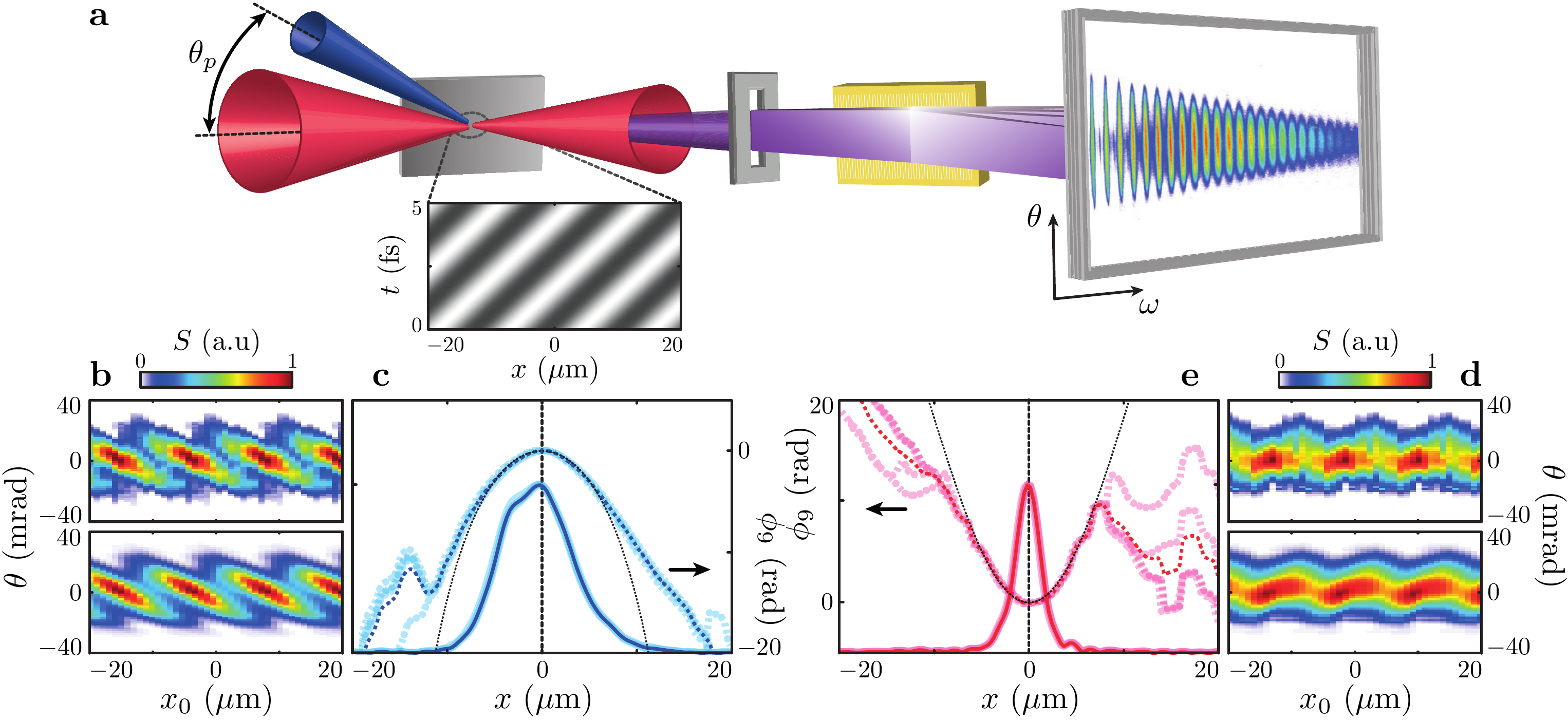}
\vskip -0.25cm 
\caption{\textbf{Applications of dynamical ptychography to attosecond pulses generated on plasma mirrors}. 
The upper panel presents the principle of the experiment, where attosecond pulses (purple cone) are generated on a plasma mirror by a spatio-temporally shaped laser field resulting from the superposition of a main beam at frequency $\omega_L$ (red cone) and a perturbation beam at frequency $2\omega_L$ (blue cone), crossing at an angle $\theta_p$ on the target. The delay $\delta t$ between the two beams is controlled on the attosecond time scale. The inset sketches the transient optical grating resulting from the superposition of the two beams at focus, for $\theta_p=35$ mrad (see SM). This is the 'object' used for the dynamical ptychography measurement, which here has a velocity $v=c/2\sin \theta_p=14.3 c$ ($c$ vacuum velocity of light). Frequency-resolved ptychographic traces are measured in the far-field using a XUV spectrometer. Two such traces are displayed in the lower row, measured in the CWE (panel b) and ROM (panel d) regimes, for harmonic 9 of the fundamental beam. The reconstructions provided by the iterative phase-retrieval algorithm are displayed below the measured traces. In both cases, the reconstruction error is of the order of 12 \%. The retrieved spatial amplitude and phase profiles of harmonic 9 in the surface plane of the plasma mirror are displayed in the central graphs (panels c and e, amplitude profile in full line, phase profile in dashed line, prediction of theoretical models \cite{leblanc2017spatial} as black dotted line). 
} 
\vskip -0.5cm
\label{fig1}
\end{figure*}

Such a fast transient optical grating can be generated by perturbing the main driving laser field, of frequency $\omega_L$, with a much weaker secondary beam of frequency $2\omega_L$, arriving at a small angle $\theta_p$ with the main beam (Fig.2a). 
At any given time $t$, the crossing of these two fields generates a sinusoidal spatial interference pattern -exactly as the object of Fig.1. Since the two waves have different frequencies, the relative phase of the two beams changes as time evolves, and the resulting interference pattern therefore drifts spatially along the plasma mirror surface - again, like in the example of Fig.1. The spatio-temporal structure of this transient optical grating can be easily calculated analytically (see SM), and is displayed in the inset of Fig.2a. Its temporal period is that of the main driving field ($T_L=2.7$ fs): the methods of attosecond metrology developed in the last 20 years have shown that such a modulation at Petahertz frequency indeed enables temporal measurements with attosecond resolution \cite{krausz2009attosecond,quere2005temporal}.

A ptychographic measurement requires scanning the position $x_0$ of the diffracting structure with respect to that of the harmonic source -imposed by the spatial profile of the main driving beam. This can be achieved with high accuracy by changing the relative delay $\delta t$ between the two beams, by small fractions of the laser optical period. At any given time $t$, this shifts the spatial interference pattern formed by their superposition, while the position of the harmonic source remains fixed. 
To ensure the required interferometric delay stability, we generate the perturbing $2\omega_L$ beam from a fraction of the main beam, thanks to an all-solid in-line optical set-up (see SM), and vary its delay with attosecond accuracy by tiny rotations of a glass plate used in transmission. Using an angularly-resolved spectrometer (Fig.2a), we then measure in a single delay scan the ptychographic trace of each individual harmonic in the spectrum of the attosecond pulse train. 

We have performed such dynamical ptychographic measurements in different interaction regimes (see SM) \cite{thaury2007plasma,Kahaly2013}, from laser intensities exceeding $10^{19}$ $W/cm^2$, where the Doppler effect described in the introduction is the main source of harmonic generation (Relativistic Oscillating Mirror regime - ROM), down to much lower intensities  of $\approx10^{17}$ $W/cm^2$, where attosecond pulses are produced by collective plasma oscillations, periodically triggered into the dense plasma by fast electron bunches (Coherent Wake Emission regime -CWE) \cite{quere2006coherent}. The comparison of the results obtained in these two regimes will provide stringent tests of the measurement method, as explained below. 

We note that different measurement schemes based on the perturbation of a laser field by its second harmonic have been demonstrated in the last decade \cite{dudovich2006measuring, kim2013manipulation, vampa2015linking}, for the characterization of attosecond pulses generated in gases or solids at laser intensities of $\sim 10^{14}$ $W/cm^2$. The justification of these schemes is based on a quantum picture, and is specific to the involved generation process. Here we provide a general conceptual framework, independent of the physics of the generation mechanism, which explains why dynamical ptychography with spatio-temporally shaped laser fields can be applied to a large variety of systems and interaction regimes -such as plasmas exposed to laser intensities $10^5$ times higher than in any previous attosecond measurement.

\section*{Spatio-temporal characterization of attosecond pulses from plasma mirrors}

Figure 2b-d displays two experimental ptychographic traces obtained for harmonic 9, in the CWE and ROM interaction regimes. These traces are then processed by a ptychographic phase retrieval algorithm (see SM), which converges to the reconstructions displayed below the measured data.

The first information obtained from the reconstruction of a single ptychographic trace is the spatial structure of the harmonic beam in the plane of the plasma mirror. The amplitude and phase profiles of harmonic 9 retrieved in the two interaction regimes are displayed in Fig.2 c-e. These spatial properties, and more particularly the phase profile, carry rich information on the physics of the harmonic generation process \cite{quere2008phase,dromey2009diffraction}, and are consistent with analytical models \cite{vincenti2014optical,malvache2013coherent} (black dotted lines in Fig.2 c-e) as well as previous spatial-only measurements \cite{leblanc2016ptychographic,leblanc2017spatial}.


\begin{figure*}[t]
\vskip -0.1cm 
\centering \includegraphics [width=\linewidth]{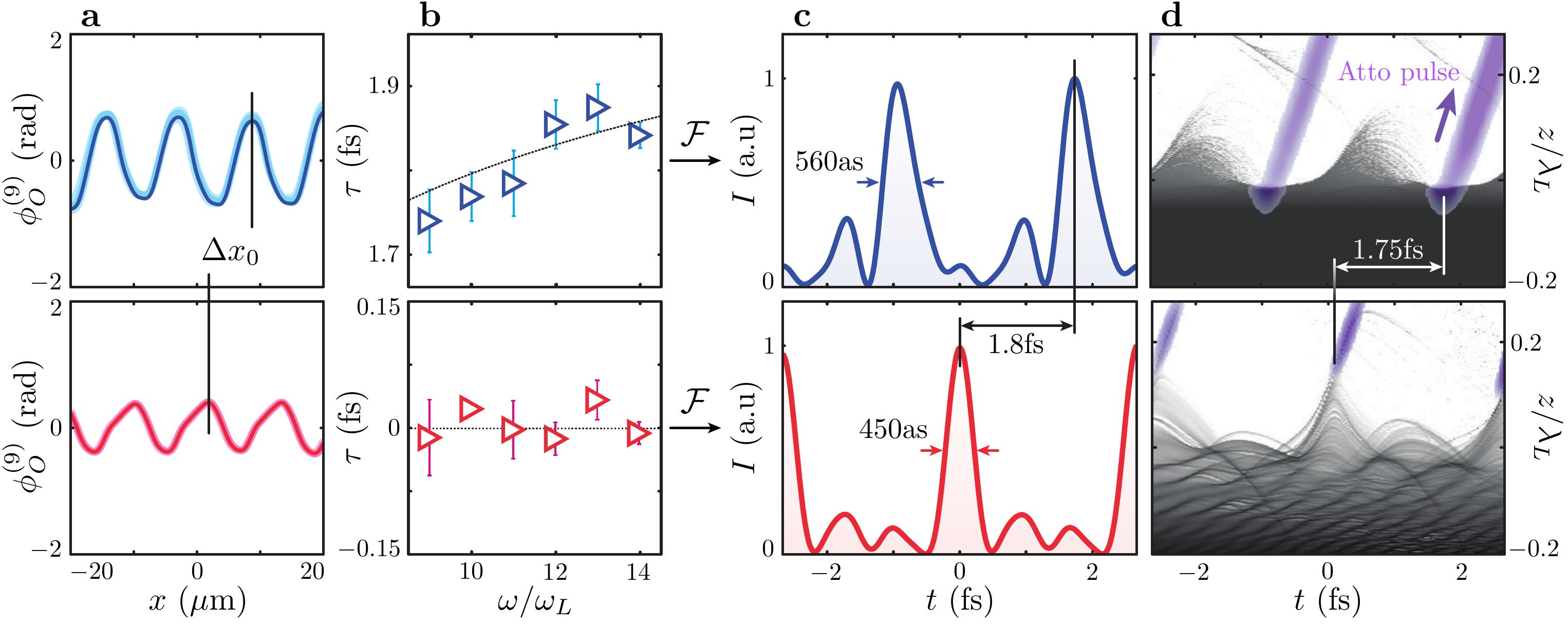}
\vskip -0.5cm 
\caption{\textbf{Temporal reconstruction of attosecond pulses from plasma mirrors}. Panel a displays the reconstruction of the spatial phase $\phi_O^{(9)}(x)$ of the transient optical grating obtained from the two ptychographic traces of Fig.2 for harmonic 9. The position of this grating encodes the emission time (group delay) of the harmonic, according to Eq.(1). By measuring the position of this grating as a function of frequency, we determine the group delay dispersion of the attosecond pulses generated in the CWE (upper row) and ROM (lower row) regimes, for harmonics 9 to 14 (panel b - black dotted lines are guides for the eye). Knowing the spectral amplitude (provided by the XUV spectrometer), we can then reconstruct the temporal profile of these attosecond pulses (panel c) by a Fourier transformation. The temporal resolution of the technique is validated by the measurement of the difference in emission times, within the laser optical period, of the CWE and ROM attosecond pulses (panel c). This measured difference matches the value determined from PIC simulations, shown in panel d that illustrates the plasma dynamics leading to attosecond pulse emission (electron density as grey color map, intensity of attosecond pulses as purple color map).}
\vskip -0.5cm
\label{fig2}
\end{figure*}

In both cases, the observed curvature of the spatial wavefront is due to the spatial variation of the laser intensity across focus. In the CWE regime, the intensity dependence of the electron dynamics at the plasma surface results in a diverging beam \cite{quere2008phase,thaury2010high}. The opposite curvature observed in the ROM regime is that of a converging harmonic beam, and has a very different physical origin. At the ultrahigh laser intensities involved in this regime, the radiation pressure exerted by the incident field on the plasma mirror surface reaches the Gigabar range, and leads to a recession of this surface typically by a few tens of nanometers \cite{dromey2009diffraction,vincenti2014optical}. Due to the varying laser intensity, this recession is stronger at the center of the focal spot than on the edges, and the plasma mirror surface thus gets dented, acquiring a parabolic shape: this is precisely the type of curved relativistic mirror that is needed to boost the reflected light intensity. 

The intensity boost that will be achieved at the plasma mirror focus is not only determined by the spatial properties of individual harmonics, obtained in Fig.2: it critically depends on the attosecond temporal compression of the reflected field. This is precisely the second type of information provided by dynamical ptychography. For each individual harmonic of frequency $\omega_n=n \omega_L$, the ptychographic algorithm provides a reconstruction of the diffracting object. According to the previous sections, the emission time $\tau(\omega_n)$ of this harmonic is encoded in the measured position offset of this object, following Eq.(1). We now experimentally validate this key feature of the measurement method.


The optical gratings retrieved from the two ptychographic traces of Fig.2 are displayed in Fig.3a. While these two measurements have been performed in different laser-plasma interaction conditions, the moving grating used for dynamical ptychography remained the same in the two cases (see SM). As expected, it is sinusoidal in space, with a spatial period $D=11.5$  $\mu m$ determined by the angle between the main and perturbing beams. More importantly, we observe that the overall position of the object retrieved for the $9^{th}$ harmonic differs by $\Delta x_0=7.5$ $\mu m$ in the two regimes. This position shift should be related to different times of emission (within the laser optical cycle) of the attosecond pulses produced in these two measurements carried out in different regimes. Converted in time using Eq.(1) with $v=14.3 c$, this corresponds to a delay $\Delta t_0= 1.8$ fs between these attosecond pulses (Fig.3c). 

Such a difference is indeed what is expected physically \cite{quere2006coherent}: in the ROM regime, attosecond pulses are emitted in the part of the optical cycle where plasma surface electrons are pulled outward by the incident laser field (Fig.3d, lower panel). Later in the laser optical cycle, these surface electrons are pushed back into the plasma, where they trigger the emission of CWE attosecond pulses (Fig.3d, upper panel). The measured delay quantitatively matches the one observed in Particle-In-Cell (PIC) simulations of the laser-plasma interaction ($\Delta t_0=1.75$ fs, Fig.3d): this provides a striking validation of the temporal sensitivity of the measurement method. 

We can now use the position offset $x_m(\omega_n)$ of the multiple objects retrieved from the ptychographic traces measured for different harmonics, to determine the group delay $\tau(\omega_n)$ of the associated attosecond pulses (see SM). These results are plotted in Fig.3b, for both the CWE and ROM regimes.
In the CWE regime, we observe that the harmonics are emitted at different times, with a total delay of $\simeq 150$ as between the emission of harmonics 9 and 14, leading to pulses slightly longer than the Fourier-transform limited duration (Fig.3c, upper panel). This group delay dispersion results from the fact that higher harmonics are emitted from denser parts of the plasma, which are located deeper into the target \cite{quere2006coherent}: they are therefore emitted later in the laser optical cycle. 
In contrast, all harmonics are found to be synchronized in the ROM regime, leading to attosecond pulses with the minimum duration allowed by the spectral bandwidth of the beam (Fig.3c, lower panel). This is because all harmonics are emitted at the same time, when the plasma mirror moves outward at relativistic velocity (Fig.3d, lower panel). 
These results constitute the first accurate temporal measurements of attosecond pulses generated from plasma mirrors.

\begin{figure}[t]
\vskip -0.1cm 
\centering \includegraphics [width=\linewidth]{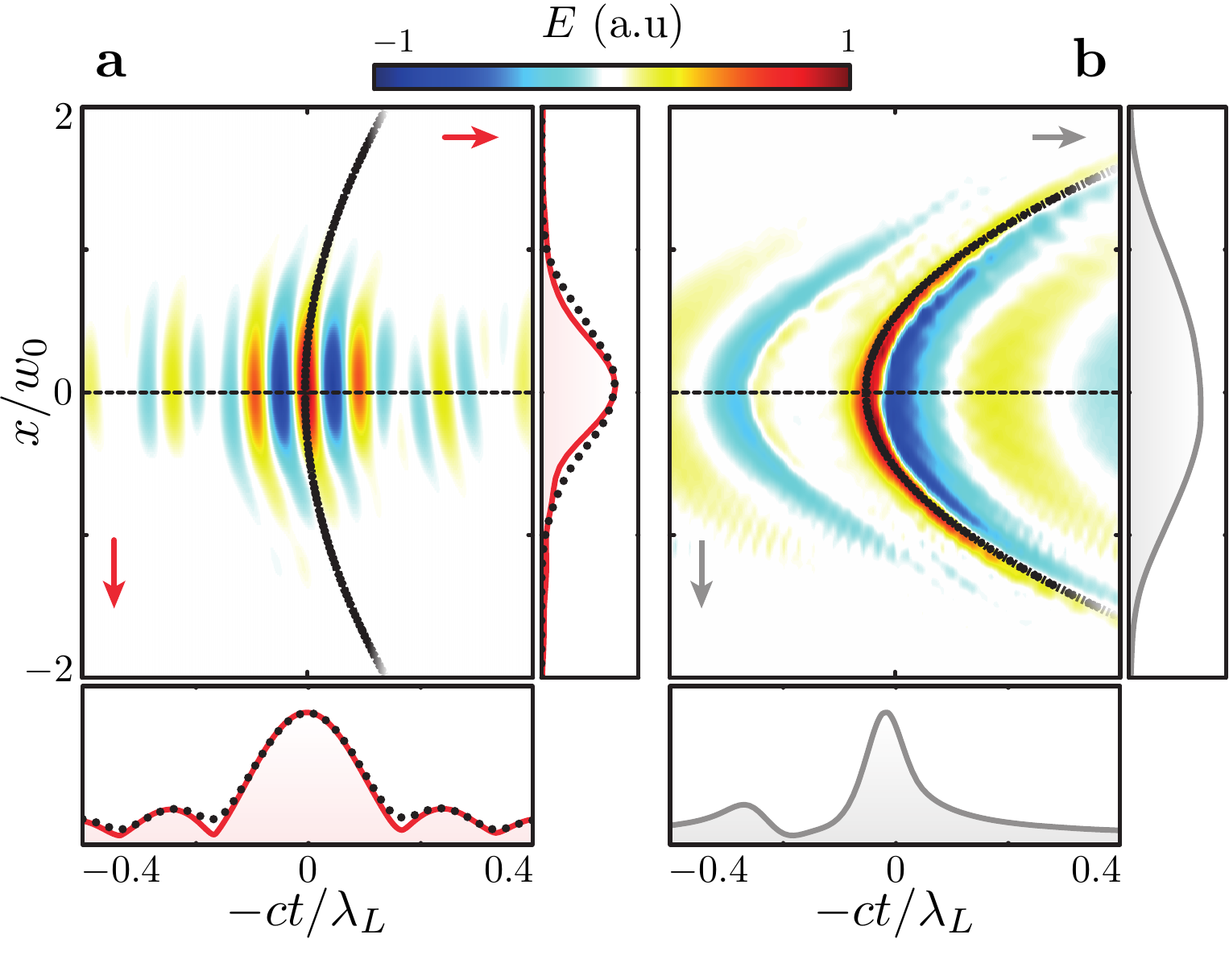}
\vskip -0.5cm 
\caption{\textbf{Spatio-temporal field of an attosecond pulse produced by a relativistic plasma mirror}. Panel a displays the experimental spatio-temporal reconstruction of the $E$-field of an attosecond pulse generated in the ROM regime, resulting from the superposition of harmonics 9 to 14, in the plane of the plasma mirror surface. Panel b displays the $E$-field now obtained from a PIC simulation where the laser intensity on target is $10^{22}$ $W/cm^2$ -nowadays achievable with a PW laser. In this second case, the displayed attosecond pulse is produced by the superposition of all generated harmonics (orders $n \geq 2$). The color lines in the bottom and side panels respectively show lineouts of the temporal profile and projections of the spatial profile in the two cases (with $w_0$ the laser beam waist diameter on target, and $\lambda_L$ the laser wavelength). In case a, the dotted black lines in these panels show results from PIC simulations carried out in the physical conditions of the experiment (see SM). The dashed black lines in the main panels a and b are guides for the eye. 
}
\vskip -0.5cm
\label{fig2}
\end{figure}

Combining the spatial and temporal information obtained from our ptychographic measurements, we can reconstruct the spatio-temporal field of the attosecond pulses formed by the superposition of harmonics 9 to 14. This field is displayed in Fig.4a for the ROM regime. Due to the spatial curvature of the wavefronts imprinted by the curved plasma mirror surface, this optimally-compressed attosecond pulse will get focused $\simeq 150$ $\mu m$ away from this surface. The combination of these temporal and spatial compressions leads to an intensity gain compared to the incident laser field. Determining the effective gain requires considering the field produced by the entire harmonic spectrum, and knowing the conversion efficiency of the laser energy to these harmonics. To this end, we use PIC simulations validated by our experimental results (Fig.4a, side and bottom panels), and estimate the intensity gain to 8 in our experiment.

\section*{Outlook: a path to the Schwinger limit}

The excellent spatio-temporal compression measured in our experiment suggests that considerably higher gains should be possible using incident laser fields with peak intensities of the order of $10^{22}$ $W/cm^2$, now available from the emerging generation of PW lasers. 
As these much higher intensities, the involved physical processes -temporal compression and tight focusing- qualitatively remain the same as the ones observed in our experiment, but their effect can be made much stronger \cite{ PhysRevLett.123.105001}. As illustrated by the simulations results of Fig.4b, shorter attosecond pulses ($\simeq 100$ $as$), associated to broader harmonic spectra, are generated with much higher conversion efficiencies, reaching $40$ \% for all harmonic orders $\geqslant 2$ in the physical conditions of Fig.4b. Due to a stronger curvature of the plasma surface, and to a larger size of the harmonic source, converging harmonic beams with larger numerical apertures are produced, which thus get focused to tighter focal spots. In the case of Fig.4b, a $300$ $nm$ focal spot is produced 60 $\mu m$ away from the plasma mirror surface, resulting in  a total intensity gain of $10^3$ that brings the light intensity in the $10^{25}$ $W/cm^2$ range.
Different approaches can be used to further enhance the focusing by these highly-controllable plasmas \cite{thaury2008coherent, wheeler2012attosecond, Monchoce2014,Leblanc2017, PhysRevLett.118.033902, yeung2017experimental} and thus make the Schwinger limit within reach \cite{PRX_bis} -such as creating the plasma mirror on a micro-fabricated curved target, using optically-shaped plasmas \cite{Monchoce2014,Leblanc2017}, or tailoring the laser wavefronts to control the harmonic beam divergence \cite{vincenti2014optical, quintard2019optics}.

In all of these cases, the advanced measurement method described here will equally apply, enabling the accurate characterization and optimization of the spatio-temporal effects induced by relativistic plasma mirrors on the reflected field. All key experimental concepts and tools are thus now available to pursue this challenging yet realistic path to the Schwinger limit based on plasma mirrors, which our results set as a new type of attosecond light source \cite{krausz2009attosecond,maroju2020attosecond} of high spatio-temporal quality.

\bibliographystyle{Science}
\bibliography{refs}

\providecommand{\noopsort}[1]{}\providecommand{\singleletter}[1]{#1}%
\begin{thebibliography}{10}

\bibitem{marklund2006nonlinear}
M.~Marklund, P.~K. Shukla, {\it Reviews of modern physics\/} {\bf 78}, 591
  (2006).

\bibitem{di2012extremely}
A.~Di~Piazza, C.~M{\"u}ller, K.~Hatsagortsyan, C.~H. Keitel, {\it Reviews of
  Modern Physics\/} {\bf 84}, 1177 (2012).

\bibitem{mourou2006optics}
G.~A. Mourou, T.~Tajima, S.~V. Bulanov, {\it Reviews of modern physics\/} {\bf
  78}, 309 (2006).

\bibitem{sauter1931verhalten}
F.~Sauter, {\it Zeitschrift f{\"u}r Physik\/} {\bf 69}, 742 (1931).

\bibitem{Heisenberg1935}
W.~Heisenberg, H.~Euler, {\it Zeitschrift f{\"u}r Physik\/} {\bf 98}, 714
  (1936).

\bibitem{schwinger1951gauge}
J.~Schwinger, {\it Physical Review\/} {\bf 82}, 664 (1951).

\bibitem{ringwald2001pair}
A.~Ringwald, {\it Physics Letters B\/} {\bf 510}, 107 (2001).

\bibitem{strickland1985compression}
D.~Strickland, G.~Mourou, {\it Optics communications\/} {\bf 55}, 447 (1985).

\bibitem{bahk2004generation}
S.-W. Bahk, {\it et~al.\/}, {\it Optics letters\/} {\bf 29}, 2837 (2004).

\bibitem{gerstner2007laser}
E.~Gerstner, {\it Nature\/} {\bf 446}, 16 (2007).

\bibitem{Yoon19}
J.~W. Yoon, {\it et~al.\/}, {\it Optics express\/} {\bf 27}, 20412 (2019).

\bibitem{landecker1952possibility}
K.~Landecker, {\it Physical Review\/} {\bf 86}, 852 (1952).

\bibitem{bulanov2003light}
S.~V. Bulanov, T.~Esirkepov, T.~Tajima, {\it Physical review letters\/} {\bf
  91}, 085001 (2003).

\bibitem{gordienko2005coherent}
S.~Gordienko, A.~Pukhov, O.~Shorokhov, T.~Baeva, {\it Physical review
  letters\/} {\bf 94}, 103903 (2005).

\bibitem{gonoskov2011ultrarelativistic}
A.~A. Gonoskov, A.~V. Korzhimanov, A.~V. Kim, M.~Marklund, A.~M. Sergeev, {\it
  Physical Review E\/} {\bf 84}, 046403 (2011).

\bibitem{PhysRevLett.123.105001}
H.~Vincenti, {\it Phys. Rev. Lett.\/} {\bf 123}, 105001 (2019).

\bibitem{solodov2006limits}
A.~Solodov, V.~Malkin, N.~Fisch, {\it Physics of plasmas\/} {\bf 13}, 093102
  (2006).

\bibitem{kapteyn1991prepulse}
H.~C. Kapteyn, M.~M. Murnane, A.~Szoke, R.~W. Falcone, {\it Optics letters\/}
  {\bf 16}, 490 (1991).

\bibitem{thaury2007plasma}
C.~Thaury, {\it et~al.\/}, {\it Nature Physics\/} {\bf 3}, 424 (2007).

\bibitem{lichters1996short}
R.~Lichters, J.~Meyer-ter Vehn, A.~Pukhov, {\it Physics of Plasmas\/} {\bf 3},
  3425 (1996).

\bibitem{thaury2010high}
C.~Thaury, F.~Qu{\'e}r{\'e}, {\it Journal of Physics B: Atomic, Molecular and
  Optical Physics\/} {\bf 43}, 213001 (2010).

\bibitem{baeva2006theory}
T.~Baeva, S.~Gordienko, A.~Pukhov, {\it Physical review E\/} {\bf 74}, 046404
  (2006).

\bibitem{dromey2006high}
B.~Dromey, {\it et~al.\/}, {\it Nature physics\/} {\bf 2}, 456 (2006).

\bibitem{krausz2009attosecond}
F.~Krausz, M.~Ivanov, {\it Reviews of Modern Physics\/} {\bf 81}, 163 (2009).

\bibitem{quere2005temporal}
F.~Qu{\'e}r{\'e}, Y.~Mairesse, J.~Itatani, {\it Journal of Modern Optics\/}
  {\bf 52}, 339 (2005).

\bibitem{orfanos2019attosecond}
I.~Orfanos, {\it et~al.\/}, {\it APL Photonics\/} {\bf 4}, 080901 (2019).

\bibitem{nomura2009attosecond}
Y.~Nomura, {\it et~al.\/}, {\it Nature Physics\/} {\bf 5}, 124 (2009).

\bibitem{quere2009attosecond}
F.~Qu{\'e}r{\'e}, {\it Nature Physics\/} {\bf 5}, 93 (2009).

\bibitem{rodenburg2008ptychography}
J.~M. Rodenburg, {\it Advances in imaging and electron physics\/} {\bf 150}, 87
  (2008).

\bibitem{thibault2008high}
P.~Thibault, {\it et~al.\/}, {\it Science\/} {\bf 321}, 379 (2008).

\bibitem{Note1}
The arrival time of frequency $\omega $ here refers to the group delay at this
  frequency, i.e. it corresponds to the temporal position of the pulse formed
  by the superposition of frequencies within a small interval centered at
  $\omega $.

\bibitem{leblanc2017spatial}
A.~Leblanc, {\it et~al.\/}, {\it Physical review letters\/} {\bf 119}, 155001
  (2017).

\bibitem{Kahaly2013}
S.~Kahaly, {\it et~al.\/}, {\it Phys. Rev. Lett.\/} {\bf 110}, 175001 (2013).

\bibitem{quere2006coherent}
F.~Qu{\'e}r{\'e}, {\it et~al.\/}, {\it Physical review letters\/} {\bf 96},
  125004 (2006).

\bibitem{dudovich2006measuring}
N.~Dudovich, {\it et~al.\/}, {\it Nature physics\/} {\bf 2}, 781 (2006).

\bibitem{kim2013manipulation}
K.~T. Kim, {\it et~al.\/}, {\it Nature Physics\/} {\bf 9}, 159 (2013).

\bibitem{vampa2015linking}
G.~Vampa, {\it et~al.\/}, {\it Nature\/} {\bf 522}, 462 (2015).

\bibitem{quere2008phase}
F.~Qu{\'e}r{\'e}, {\it et~al.\/}, {\it Physical review letters\/} {\bf 100},
  095004 (2008).

\bibitem{dromey2009diffraction}
B.~Dromey, {\it et~al.\/}, {\it Nature Physics\/} {\bf 5}, 146 (2009).

\bibitem{vincenti2014optical}
H.~Vincenti, {\it et~al.\/}, {\it Nature communications\/} {\bf 5}, 3403
  (2014).

\bibitem{malvache2013coherent}
A.~Malvache, A.~Borot, F.~Qu{\'e}r{\'e}, R.~Lopez-Martens, {\it Physical Review
  E\/} {\bf 87}, 035101 (2013).

\bibitem{leblanc2016ptychographic}
A.~Leblanc, S.~Monchoc{\'e}, C.~Bourassin-Bouchet, S.~Kahaly, F.~Qu{\'e}r{\'e},
  {\it Nature Physics\/} {\bf 12}, 301 (2016).

\bibitem{thaury2008coherent}
C.~Thaury, {\it et~al.\/}, {\it Nature physics\/} {\bf 4}, 631 (2008).

\bibitem{wheeler2012attosecond}
J.~A. Wheeler, {\it et~al.\/}, {\it Nature Photonics\/} {\bf 6}, 829 (2012).

\bibitem{Monchoce2014}
S.~Monchoc\'e, {\it et~al.\/}, {\it Phys. Rev. Lett.\/} {\bf 112}, 145008
  (2014).

\bibitem{Leblanc2017}
A.~Leblanc, {\it et~al.\/}, {\it Nature Physics\/} {\bf 13}, 440 EP  (2017).

\bibitem{PhysRevLett.118.033902}
A.~Denoeud, L.~Chopineau, A.~Leblanc, F.~Qu\'er\'e, {\it Phys. Rev. Lett.\/}
  {\bf 118}, 033902 (2017).

\bibitem{yeung2017experimental}
M.~Yeung, {\it et~al.\/}, {\it Nature Photonics\/} {\bf 11}, 32 (2017).

\bibitem{PRX_bis}
H.~Vincenti, F.~Qu{\'e}r{\'e}, {\it submitted to Physical Review X\/}  (2020).

\bibitem{quintard2019optics}
L.~Quintard, {\it et~al.\/}, {\it Science advances\/} {\bf 5}, eaau7175 (2019).

\bibitem{maroju2020attosecond}
P.~K. Maroju, {\it et~al.\/}, {\it Nature\/} {\bf 578}, 386 (2020).

\end{thebibliography}

\section*{Acknowledgments}
The datasets generated during and/or analysed during the current study are available from the corresponding author on reasonable request. The WARP code is publicly available at \url{https://bitbucket.org/berkeleylab/warp/src/master/}.  The PICSAR library is publicly available at \url{https://bitbucket.org/berkeleylab/picsar/src/master/}. 
We thank F. R\'eau, C. Pothier, and D. Garzella for operating the UHI100 laser. The research received financial support from the European Research Council (ERC
Grant Agreement No. 694596), from LASERLAB-EUROPE (Grant Agreement No. 871124, European Union Horizon 2020 research and innovation programme),  from Investissements d'Avenir LabEx PALM (ANR-10-LABX-0039-PALM) and from Agence Nationale de la Recherche (ANR-18-ERC2-0002). An award of computer time was provided by the INCITE program (project 'PlasmInSilico'). This research used resources of the Argonne Leadership Computing Facility, which is a DOE Office of Science User Facility supported under Contract DE-AC02-06CH11357. 


\end{document}